\documentclass[12pt]{article}

\usepackage {graphicx}
\usepackage{amssymb}
\usepackage{amsmath}
\usepackage{amsbsy}
\usepackage{bm}
\usepackage{wasysym}
\usepackage{textcomp}
\usepackage {graphicx}
\usepackage[export]{adjustbox}
\usepackage[]{algorithm2e}
\usepackage[titletoc,title]{appendix}
\usepackage{setspace}

\usepackage[usenames]{xcolor}

\RequirePackage[colorlinks,citecolor=blue,urlcolor=blue]{hyperref}

\newtheorem{theorem}{Theorem}
\newtheorem{corollary}{Corollary}
\newtheorem{proposition}{Proposition}
\newtheorem{lemma}{Lemma}
\newtheorem{example}{Example}

\newtheorem{definition}{Definition}

\newcommand{\beq}{\begin{equation}}
\newcommand{\eeq}{\end{equation}}
\newcommand{\beas}{\begin{eqnarray*}}
\newcommand{\eeas}{\end{eqnarray*}}
\newcommand{\bea}{\begin{eqnarray}}
\newcommand{\eea}{\end{eqnarray}}
\newcommand{\bei}{\begin{itemize}}
\newcommand{\eei}{\end{itemize}}
\newcommand{\ben}{\begin{enumerate}}
\newcommand{\een}{\end{enumerate}}
\newcommand{\bet}{\begin{theorem}}
\newcommand{\eet}{\end{theorem}}
\newcommand{\bel}{\begin{lemma}}
\newcommand{\eel}{\end{lemma}}
\newcommand{\bep}{\begin{proposition}}
\newcommand{\eep}{\end{proposition}}
\newcommand{\bed}{\begin{definition}}
\newcommand{\eed}{\end{definition}}
\newcommand{\bec}{\begin{corollary}}
\newcommand{\eec}{\end{corollary}}
\newcommand{\bex}{\begin{example}}
\newcommand{\eex}{\end{example}}

\def\0{\boldsymbol{0}}

\def\x{\boldsymbol{x}}

\begin{document}

\title{Optimal design for high-throughput screening via false discovery rate control}
\author{Tao Feng$^1$, Pallavi Basu$^2$, Wenguang Sun$^3$, Hsun Teresa Ku$^4$,  Wendy J. Mack$^1$}
\date{}

\footnotetext[1]{Department of Preventive Medicine, Keck School of Medicine, University of Southern California.}
\footnotetext[2]{Department of Statistics and Operations Research, Tel Aviv University.}
\footnotetext[3]{Department of Data Sciences and Operations, University of Southern California. The research of Wenguang Sun was supported in part by NSF grant DMS-CAREER 1255406.} 
\footnotetext[4]{Department of Diabetes and Metabolic Diseases Research, City of Hope National Medical Center.}
\maketitle

\begin{abstract}

High-throughput screening (HTS) is a large-scale hierarchical process in which a large number of chemicals are tested in multiple stages. Conventional statistical analyses of HTS studies often suffer from high testing error rates and soaring costs in large-scale settings. This article develops new methodologies for false discovery rate control and optimal design in HTS studies. We propose a two-stage procedure that determines the optimal numbers of replicates at different screening stages while simultaneously controlling the false discovery rate in the confirmatory stage subject to a constraint on the total budget. The merits of the proposed methods are illustrated using
both simulated and real data. We show that the proposed screening procedure effectively controls the error rate and the design leads to improved detection power. This is achieved at the expense of a limited budget.
\end{abstract}

\thispagestyle{empty}


\section{Introduction}

In both pharmaceutical industries and academic institutions, high throughput screening (HTS) is a primary approach for selection of biologically active agents or drug candidates from a large number of compound \cite{Ma06}. Over the last 20 years, HTS has played a crucial role in fast-advancing fields such as screening of small molecule or short interference RNA (siRNA) screening in stem cell biology \cite{Xu08} or molecular biology (\cite{Mo06} \cite{Ec06}), respectively. HTS is a large-scale and multi-stage process in which the number of investigated compounds can vary from hundreds to millions. The stages involved are target identification, assay development, primary screening, confirmatory screening, and follow-up of hits \cite{Go}.

The accurate selection of useful compounds is an important issue at each aforementioned stage of HTS. In this article we focus on statistical methods at the primary and confirmatory screening stages. A library of compounds is first tested at the primary screening stage, generating an initial list of selected positive compounds, or `hits'. 
The hits are further investigated at the confirmatory stage, generating a list of `confirmed hits'. In contrast to the assays used at the primary screening stage, the assays for the confirmatory screening stage are more accurate but more costly. 

Conventional statistical methods, such as the z-score, robust z-score, quartile-based, and strictly standardized mean difference methods have been used for selection of compounds as hits or confirmed hits (\cite{Ma06} \cite{Go}). However, these methods are highly inefficient due to three major issues. First, these methods ignore the multiple comparison problem. When a large number of compounds are tested simultaneously, the inflation of type I errors or false positives becomes a serious issue and may lead to large financial losses in the follow-up stages. The control of type I errors is especially crucial at the confirmatory screening stage due to the higher costs in the hits follow-up stage. The family-wise error rate (FWER), the probability of making at least one type I error, is often used to control the multiplicity of errors (\cite{We93} \cite{Du02} \cite{Sh95} \cite{Ho79} \cite{Ho88}). However, in large-scale HTS studies, the FWER criterion becomes excessively conservative, such that it fails to identify most useful compounds. In this study we consider a more cost-effective and powerful framework for large-scale inference; the goal is to control the false discovery rate (FDR) \cite{Be95}, the expected proportion of false positive discoveries among all confirmed hits.

Second, the data collected in conventional HTS studies often have very low signal to noise ratios (SNR). For example, in most HTS analyses, only one measurement is obtained for each compound at the primary screening stage \cite{Ma06}; existing analytical strategies often lead to a high false negative rate, an overall inefficient design, and hence inevitable financial losses (since missed findings will not be pursued). 

Finally, in the current HTS designs, an optimal budget allocation between the primary screening and confirmatory screening stages is not considered. Ideally the budgets should be allocated efficiently and dynamically to maximize the statistical power. Together, the overwhelming number of targets in modern HTS studies and the lack of powerful analytical tools have contributed to high decision error rates, soaring costs in clinical testing and declining drug approval rates \cite{Do03}.

This article proposes a new approach for the design and analysis of HTS experiments to address the aforementioned issues. We first formulate the HTS design problem as a constrained optimization problem where the goal is to maximize the expected number of true discoveries subject to the constraints on the FDR and study budget. We then develop a simulation-based computational procedure that dynamically allocates the study budgets between the two stages and effectively controls the FDR in the confirmatory stage. Simulation studies are conducted to show that, within the same study budget, the proposed method controls the FDR effectively and identifies more useful compounds compared to conventional methods. Finally, we confirm the usefulness of our methods employing the data obtained from a chemical screening.

Powerful strategies and methodologies have been developed for the design and analysis of multistage experiments. However, these existing methods cannot be directly applied to the analysis of HTS data. Satagopan et al.~\cite{Sa04} proposed a two-stage design for genome-wide association studies; compared to conventional single-stage designs, their two-stage design substantially reduces the study cost, while maintaining statistical power. However, the error control issue and optimal budget allocation between the stages were not considered. Posch et al.~\cite{Ze08} developed an optimized multistage design for both FDR and FWER control in the context of genetic studies. The above methods are not suitable for HTS studies since the varied cost per compound at different stages were not taken into account. M\"{u}ller et al.~\cite{Mu04} and Rossell and M\"{u}ller~\cite{Ro13} studied the optimal sample size problem and developed a two-stage simulation based design in a decision theoretical framework with various utility functions. However, it is unclear how the sample size problem and budget constraints can be integrated into a single design. In addition, the varied stage-wise costs were not considered in their studies. Other related works on multiple comparison issue in multistage testing problems include Dmitrienko et al.~\cite{Di07} and Goeman and Mansmann \cite{Go08}. The results cannot be applied to our problem due to similar aforementioned issues. Compared to existing methods, our data-driven procedure simultaneously addresses the error rate control, varied measurement costs across stages and optimal budget allocation, and is in particular suitable for HTS studies.

The remainder of the article is organized as follows. Section \ref{sec_model} presents the model, problem formulation, and proposed methodology. Numerical results are given in Section \ref{sec_num}, where we first compare the proposed method with the conventional methods using simulations, and then illustrate the method using HTS data. Section \ref{sec_disc} concludes the article with a discussion of results and future work. Technical details of the computation are provided in  Appendix \ref{sec_app}.

\section{Model, Problem Formulation, and Methods} \label{sec_model}

We first introduce a multi-stage two-component random mixture model for HTS data (Section \ref{random mixture}), and then formulate the question of interest as a constrained optimization problem (Section \ref{constraint optimization}). Finally, we develop a simulation-based computational algorithm for optimal design and error control in HTS studies (Sections \ref{fdr control} and \ref{data-driven}). 

\subsection{A random effect multi-stage normal mixture model}\label{random mixture}

We start with a popular random mixture model for single-stage studies and then discuss how it may be generalized to describe HTS data collected over multiple stages. All $m$ compounds in the HTS library can be divided into two groups: null cases (noises) and non-null cases (useful compounds). Let $p$ be the proportion of non-nulls and $\theta_i$ be a Bernoulli($p$) variable, which takes the value of 0 for a null case and 1 for a non-null case. In a two-component random mixture model, the observed measurements $x_i$ are assumed to follow the conditional distribution
\begin{equation} \label{rmm}
x_i|\theta_i \sim (1-\theta_i) f_0 + \theta_i f_1,
\end{equation}
for $i=1, \cdots, m$, where $f_0$ and $f_1$ are the null and non-null density functions respectively. Marginally, we have
\begin{equation} \label{sim model}
x_i \sim f := (1-p) f_0 + p f_1.
\end{equation} 
The marginal density $f$ is also referred to as the mixture density. 

The single stage random mixture model can be extended to a two-stage random mixture model to describe the HTS. Specifically, let $x_{1i}$ denote the observed measurement for the $i$th compound at stage I (i.e.~the primary screening stage). Denote by $p_1$ the proportion of useful compounds at stage I. The true state of nature of compound $i$ is denoted by $\theta_{1i}$, which is assumed to be a Bernoulli($p_1$) variable. Correspondingly, let $x_{2j}$ and $\theta_{2j}$ denote the observed measurement and true state of nature for the $j$th compound at stage II (i.e.~the confirmatory screening stage) respectively. 

At stage I the observed measurements are denoted by $\pmb x_1 = (x_{11}, \cdots, x_{1m_1 })$, where $m_1$ is the number of compounds tested at stage I. According to (\ref{rmm}), we assume the following random mixture model:
\[
x_{1i} |\theta_{1i} \sim (1 - \theta_{1i}) f_{10} + \theta_{1i} f_{11},               
\]                                         
where $f_{10}$ and $f_{11}$ are assumed to be the stage I density functions for the null and non-null cases respectively. We shall first focus on a normal mixture model and take both $f_{10}$ and $f_{11}$ as normal density functions. Extensions to more general situations are considered in the discussion section.

First consider the situation where there is only one replicate per compound. Marginally the means of both the null and non-null densities are assumed to be zero. The assumption of zero mean can be more generalized with more technicalities. Here a zero mean can indicate that the non-null can be either a higher or a lower valued observation. Initial transformation of the data may be needed to justify this.

We consider a hierarchical model. Let the measurement `noise' distribution follow $N(0, \sigma_{10}^2)$. The observations are viewed as $x_{1i} = \mu_{1i} + \varepsilon_{1i}$ where $\varepsilon \sim N(0, \sigma_{10}^2)$ and the signals are distributed as $\mu_{1i} \sim N(0, \sigma_{1\mu}^2)$. Then $\sigma_{10}$ denotes the standard deviation of the null distribution $f_{10}$, and let $\sigma_{11}$ denote the standard deviation of the non-null distribution $f_{11}$, where 
\[
\sigma_{11}^2 = \sigma_{1\mu}^2 + \sigma_{10}^2.
\]
If there are $r_1$ replicates per compound, then $f_{10}$ is a normal density function with a rescaled standard deviation of $\sigma_{10}/\sqrt{r_1}$; $f_{11}$ is a normal density function with adjusted standard deviation $(\sigma_{1\mu}^2 + \sigma_{10}^2/r_1)^{1/2}$. 

Suppose that $m_2$ compounds are selected from stage I to enter stage II. The observed measurements, denoted as $x_2 = (x_{21}, \cdots, x_{2m_2})$, are assumed to follow another random mixture model:
\begin{equation}\label{sim proc}
x_{2j}|\theta_{2j} \sim (1 - \theta_{2j}) f_{20} + \theta_{2j} f_{21}.
\end{equation}
Here $f_{20}$ and $f_{21}$ are the null and non-null density functions respectively, and we assume normality for both densities. Consider the situation with one replicate per hit. Here again, marginally, the mean of both the null and non-null densities are assumed to be zero. Let $\sigma_{20}$ be the standard deviation of the distribution of the null cases, and $\sigma_{21}$ the standard deviation of the distribution of the non-null cases with $\sigma_{21}^2 = \sigma_{2\mu}^2 + \sigma_{20}^2$, where $\sigma_{2\mu}$ denotes the standard deviation of the signals. For $r_2$ replicates per hit, $f_{20}$ is a normal density function with a rescaled standard deviation of $\sigma_{20}/\sqrt{r_2}$; and $f_{21}$ is a normal density function with adjusted standard deviation $(\sigma_{2\mu}^2 + \sigma_{20}^2/r_2)^{1/2}$.

\subsection{Problem formulation}\label{constraint optimization}

We aim to find the most efficient HTS design that identifies the largest number of true confirmed hits subject to the constraints on both the FDR and available funding. Our design provides practical guidelines on the choice of the optimal number of replicates at stage I, the selection of the optimal number of hits from stage I, and the optimal number of replicates at stage II accordingly, determined by study budget. 

This section formulates a constrained optimization framework for the two-stage HTS analysis. We start by introducing some notation. Let $B$ denote the total available budget. The budgets for stage I and stage II are denoted by $B_1$ and $B_2$ respectively. At stage I, let $m_1$ denote the number of compounds to be screened, $r_1$ the number of replicates per compound (same $r_1$ for every compound), $c_1$ the cost per replicate, and $A_1$ the subset of hits that are selected by stage I to enter to stage II. At stage II, let $r_2$ be the number of replicates per hit (same $r_2$ for every hit), $c_2$ the cost per replicate, and $A_2$ the subset of final confirmed hits. We use $|A_1|$ and $|A_2|$ to denote the cardinalities of the compound sets. The relations of variables above can be described by the following equations:
\begin{equation}\label{budget1}
B = B_1 + B_2,
\end{equation}
where,
\begin{equation}                         
B_1 = c_1 r_1 m_1,
\end{equation}
and
\begin{equation}\label{budget3}
B_2 = c_2 r_2 |A_1|.                                                                
\end{equation}
The optimal design involves determining the optimal combination of $r_1$ and $|A_1|$ to maximize the expected number of confirmed hits subject to the constraints on the total budget and the FDR at stage II. 

It is important to note that the two-stage study can be essentially viewed as a screen-and-clean design, and we only aim to clean the false discoveries at stage II due to the high cost at the subsequent hits follow-up stage. The main purpose of stage I is to reduce the number of compounds to be investigated in the next stage to save study budget; the control of false positive errors is not a major concern at stage I. 

In the next two subsections, we first review the methodology on FDR control and then develop a data-driven procedure for analysis of two-stage HTS experiments. 

\subsection{FDR controlling methodology in a single-stage random mixture model}\label{fdr control}

Due to the high cost of hits follow-up, we propose to control the FDR at the confirmatory screening stage (stage II). The FDR is defined as the expected proportion of false positive discoveries among all rejections, where the proportion is zero if no rejection is made. In HTS studies, the number of compounds to be screened can vary from hundreds to millions and conventional methods for controlling the FWER are overly conservative. Controlling the FDR provides a more cost-effective framework in large-scale testing problems and has been widely used in various scientific areas such as bioinformatics, proteomics, and neuroimaging.

Next we briefly describe the $z$-value adaptive procedure proposed in Sun and Cai \cite{Su07}. In contrast to the popular FDR procedures that are based on p-values, the method is based on thresholding the local false discovery rate (Lfdr) \cite{Ef01}. The Lfdr is defined as 
\[
\mbox{Lfdr}(x) := P(\mbox{null}|x \mbox{  is observed}), 
\]
the posterior probability that a hypothesis is null given the observed data. In a compound decision theoretical framework, it was shown by Sun and Cai \cite{Su07} that the Lfdr procedure outperforms p-value based FDR procedures. The main advantage of the method is that it efficiently pools information from different samples. The Lfdr procedure is in particular suitable for our analysis because it can be easily implemented in the two-stage computational framework that we have developed. As we shall see, the Lfdr statistics (or the posterior probabilities) can be computed directly using the samples generated from our computational algorithms.   

Consider a random mixture model (\ref{rmm}). The Lfdr can also be computed as
\[
\mbox{Lfdr}_i= \frac{(1 - p) f_0 (x_i)}{f(x_i)},
\]
where $x_i$ is the observed data associated with compound $i$, and
\[
f(x_i) := (1 - p) f_0 (x_i) + p f_1 (x_i)
\]
is the marginal density function. The Lfdr procedure operates in two steps: ranking and thresholding. In the first step, we order the Lfdr values from the most significant to the least significant: $\mbox{Lfdr}_{(1)} \leq \mbox{Lfdr}_{(2)} \leq \cdots \leq \mbox{Lfdr}_{(m)}$, the corresponding hypotheses are denoted by $H_{(i)}, i = 1, \cdots ,m$. In the second step, we use the following step-up procedure to determine the optimal threshold: Let
\begin{equation}\label{lfdr proc}
k = \max \left \{j: \frac{1}{j} \sum_{i=1}^j \mbox{Lfdr}_{(i)} \leq \alpha \right \}.
\end{equation}
Then reject all $H_{(i)}, i = 1, \cdots, k$. This procedure will be implemented in our design to control the FDR at stage II. When the distributional informations (for e.g., the non-null proportion $p$ and the null and non-null densities) are unknown, we need to estimate it from the data.  Related estimation issues, as well as the proposed two-stage computational algorithms, are discussed in the next subsection.

\subsection{Data-driven computational algorithm}\label{data-driven}

This section proposes a simulation-based computational algorithm that controls the FDR and dynamically allocates the study budgets between the two stages. The main ideas are described as follows. For each combination of specific values of $r_1$ and $|A_1|$, we apply the Lfdr procedure and estimate $E |A_2|$, the expected size of $A_2$ (recall that $A_2$ is the subset of final confirmed hits). The optimal design then corresponds to the combination of $r_1$ and $|A_1|$ that yields the largest expected size of confirmed hits, subject to the constraints on the FDR and study budget. A detailed description of our computational algorithm for a two-point normal mixture model is provided in Appendix \ref{sec_app_2}. The model will be used in both our simulation studies and real data analysis. 

The key steps in our algorithms include (i) estimating the Lfdr statistics, and (ii) computing the expected sizes of confirmed hits via simulations. The algorithm can be extended to a $k$-point normal mixture without essential difficulty by, for example, implementing the estimation methods described by Kom\'{a}rek \cite{Ko09}. 

Suppose that, prior to stage I we have obtained information about the unknown parameters $\sigma_{10}, \sigma_{11}$, and $p_1$ from some pilot studies. If not, we must obtain at least one replicate of $x_1 = (x_{11}, \cdots ,x_{1m_1})$ to proceed. Then by using the MC (Monte Carlo) based algorithm described in the Appendix \ref{sec_app_1}, we can estimate the unknown parameters $\sigma_{10}, \sigma_{11}$, and $p_1$. Next we simulate measurements for a given value of $r_1$ according to model (\ref{sim model}). We then select $|A_1|$ most significant compounds to proceed to stage II. Different combinations of $r_1$ and $|A_1|$ will be considered to optimize the design. At stage II, we again have two situations of which the ideal case is that we have some prior knowledge about the values of $\sigma_{20}$ and $\sigma_{21}$ from some pilot studies. Otherwise we need to obtain some preliminary data $x_2 = (x_{21}, \cdots ,x_{2m_2})$ with at least one replicate. Using MC algorithm, we can estimate the unknown parameters $\sigma_{20}$ and $\sigma_{21}$. Following (\ref{sim proc}) and information on $\theta_{1i}$ from stage I, we simulate measurements for a specific value of $r_2$ calculated by the relations (\ref{budget1})--(\ref{budget3}). Applying the Lfdr procedure (\ref{lfdr proc}) at the nominal FDR level, we can determine the subset of confirmed hits $A_2$.

For each combination of $r_1$ and $|A_1|$, the algorithm will be repeated $N$ times (in our numerical studies and data analysis we use $N=100$); the expected size of confirmed hits can be calculated as the average sizes of these subsets. Therefore we can obtain the expected sizes of confirmed hits for all combinations of $r_1$ and $|A_1|$. Finally the optimal design is determined as the combination of $r_1$ and $|A_1|$ that yields the largest expected size of confirmed hits.

The operation of our computational algorithm implies that the FDR constraint will be fulfilled and the detection power will be maximized. 

\section{Numerical Results} \label{sec_num}
We now turn to the numerical performance of our proposed method via simulation studies and a real data application. In the simulation studies, we compare the FDR and the number of identified true positive compounds of the proposed procedure with replicated one-stage and two-stages Benjamini-Hochberg (BH) procedures. The methods are compared for efficiency at the same study budget.
To investigate the numerical properties of the proposed method in different scenarios, we consider two simulation settings, discussed in Section \ref{sim}. The real data application is discussed in Section \ref{sec_real}.

\subsection{Simulation studies} \label{sim}
First we describe the two procedures that we compare our proposed procedure with. Both the methods have the same (after rounding) total budget constraint as ours. Stage I replicated BH procedure: $r_1$ replicates of stage I observations for all the $m_1$ compounds with $r_1 = \lceil B/c_1m_1 \rceil$ are obtained, where $\lceil x \rceil$ denotes the nearest integer greater than or equal to $x$. The p-values under the null model, taking into account the replication, are obtained and the BH procedure is applied to identify significant compounds. Two-stages replicated BH procedure: first 10 replicates of stage I observations for all the $m_1$ compounds are obtained. Further compounds with z-scores larger than two standard deviations in absolute value are screened for stage II with the maximum possible replicates that fits the budget of the study. At stage II the BH procedure is used to determine the final confirmed hits.

The two-component random mixture model (\ref{rmm}) is used to simulate data for both stages, which were used as the preliminary data. We then followed the procedure described in Section \ref{data-driven} with $N= 100$. For the conventional methods, we followed the approaches described above. The simulated true state of nature of compound $i$, that is $\theta_{1i}$, is used to determine the number of true positive compounds and FDR. All results are reported as average over 200 replications.

The number of compounds at stage 1 ($m_1$) is taken to be 500. The total budget ($B$) was \$250,000. The cost per replicate at stage I ($c_1$) is chosen to be \$20 whereas that for stage II ($c_2$) is chosen to be \$50 with stage II experiments assumed to be three times more precise (one over variance). The error variance ($\sigma_{10}$) is taken to be 1. The variance for the signals ($\sigma_{11}$) is taken to be 6.25. To compare the various methods we plot the realized FDR. We also plot the expected number of true positives (ETP) among the final confirmed hits as an indication for efficiency or optimality.

For our proposed method, a series of $r_1$ taking the values of $1, 2, \cdots, 25$ was tested. For every such value of $r_1$, the choice of $|A_1|$ is varied from $1, 2, \cdots \min\{(B - c_1r_1m_1)/c_2, m_1\}$ where the last number is rounded to the lower integer. Further given a choice of $r_1$ and $|A_1|$, the number of replicates at stage II ($r_2$) is computed as $\lfloor (B - c_1r_1m_1)/c_2|A_1| \rfloor$, where $\lfloor x \rfloor$ indicates the integer lower than or equal to $x$.

\subsubsection{Simulation study 1}\label{section3.1}
\begin{figure}
\centering
\includegraphics[max size={\textwidth}{\textheight}]{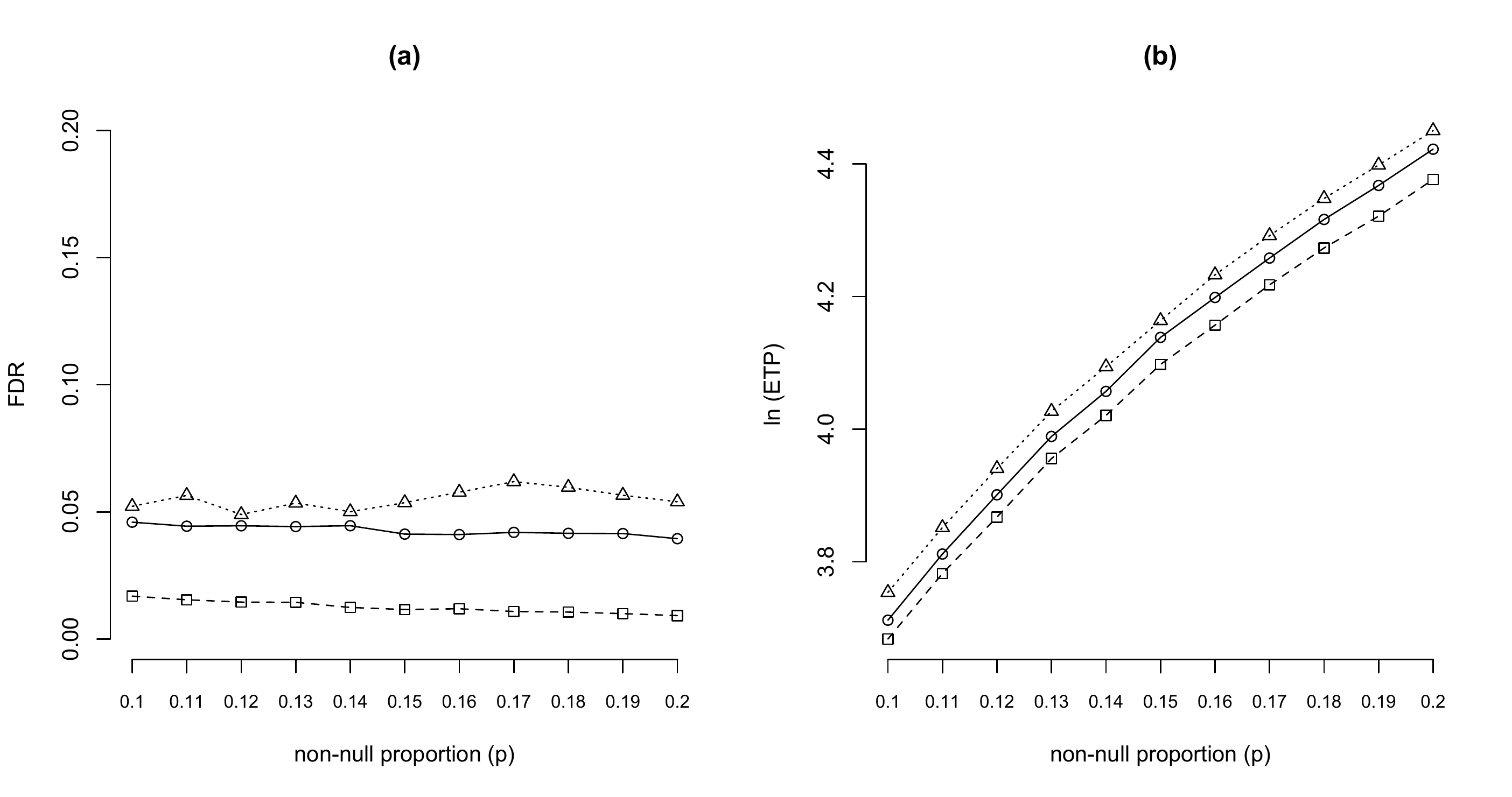} 
\caption{Realized FDR and ln of ETP are computed over 200 replications for simulation study 1. $\bigtriangleup$ is proposed method, $\bigcirc$ is two-stages BH, and $\Box$ is one-stage BH. All methods have a budget of about \$250,000.}
\label{figure1_simulations}
\end{figure}

A series of proportions $p_1$ of positive compounds at $0.1, 0.11, \cdots, 0.20$ is used to simulate the data. The results are summarized in Figure \ref{figure1_simulations}. The following observations can be made based on the results from this simulation study. First, from plot (a), all the methods control the FDR at the desired 0.05 level. The one-stage BH is overly conservative. The two-stages BH nearly controls the FDR at 0.05. Second, from plot (b), the proposed method is more powerful than the two competitive methods, and in particular has more power than the two-stages BH method. This validates our belief that choosing a fixed number of replications is suboptimal in many situations. This also indicates that conventional z-score based methods can be suboptimal as well.

\subsubsection{Simulation study 2}\label{section3.2}
\begin{figure}
\centering
\includegraphics[max size={\textwidth}{\textheight}]{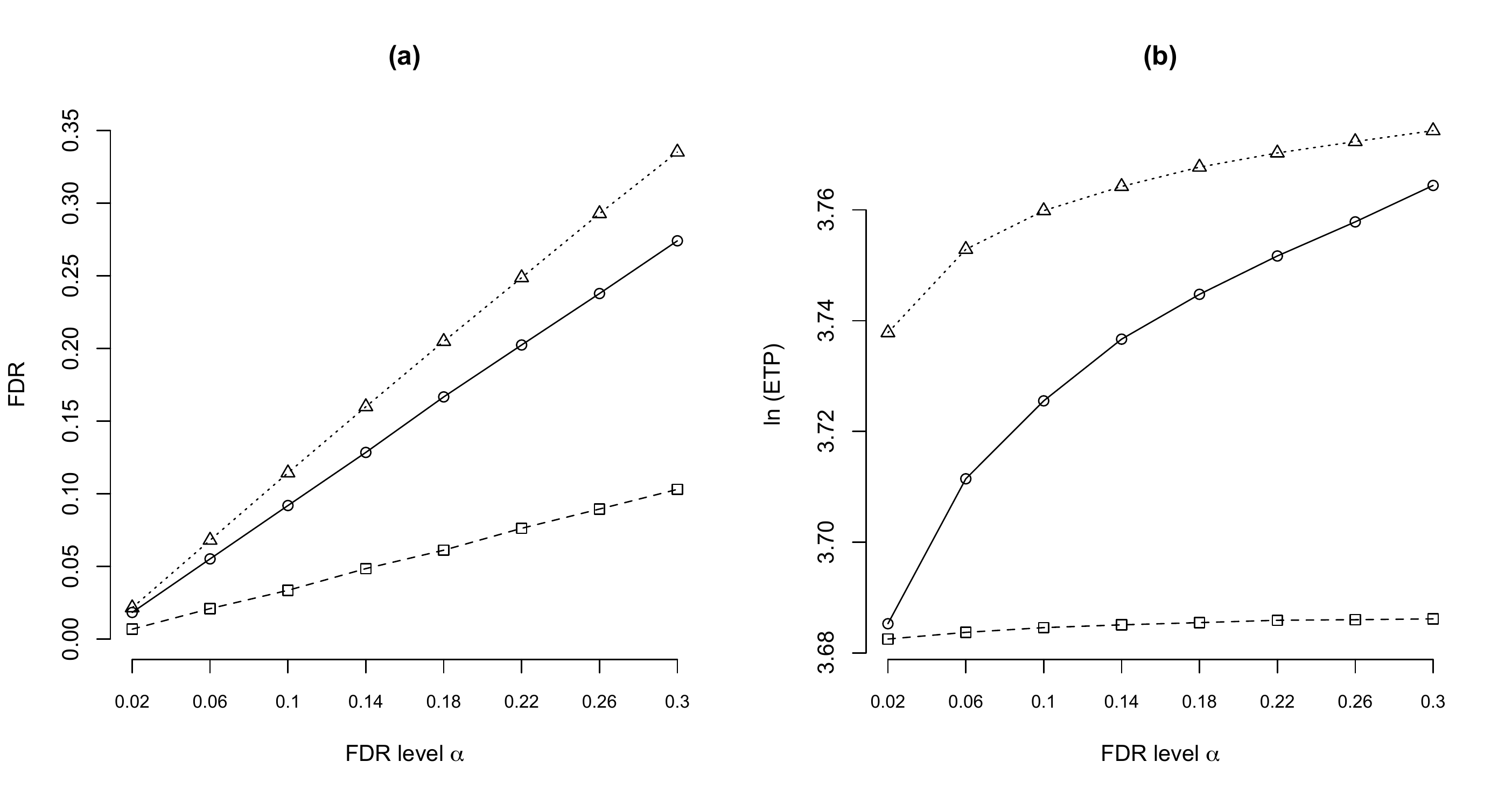} 
\caption{Realized FDR and ln of ETP are computed over 200 replications for simulation study 2. $\bigtriangleup$ is proposed method, $\bigcirc$ is two-stages BH, and $\Box$ is one-stage BH. All methods have a budget of about \$250,000.}
\label{figure2_simulations}
\end{figure}

In this simulation, we used only a single value of $p_1 = 0.1$ to simulate the data. The FDR level is varied over a range of values from 0.02--0.3. Other simulation parameters are the same as that in simulation study 1. The results are summarized in Figure \ref{figure2_simulations}. The following additional observations can be made. The proposed and two-stages BH tries to adapt to the increasing FDR level, indicated by the slope close to 1 in the plot (a). From plot (b) we observe that the gain in power is higher when the desired level of FDR is lower. This can be explained by higher number of rejections (or higher number of confirmed hits) while maintaining the FDR, which is a ratio. This neatly ties to the fact that our proposed procedure exploits the number of confirmed hits as the objective value to maximize.

\newpage

\subsection{Application to a small molecule HTS study} \label{sec_real}
\begin{figure}
\centering
\includegraphics[max size={\textwidth}{\textheight}]{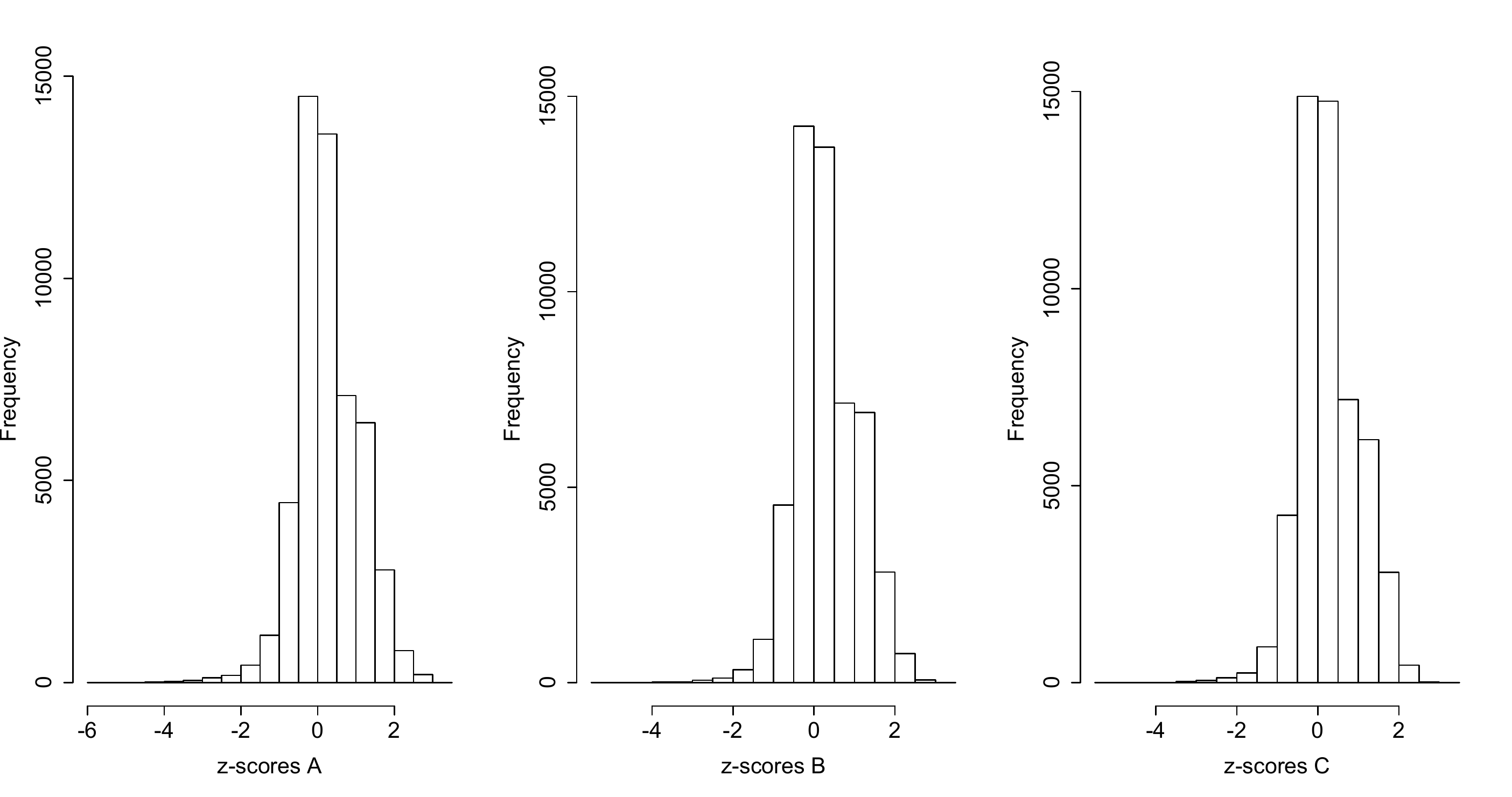} 
\caption{Histogram of z-scores A, B, and C. The plot on the left displays skewness towards negative values.}
\label{figure1_data}
\end{figure}

We analyze the study in Mckoy et al.~\cite{Mc12}. The goal of the study is to identify new inhibitors of the amyloid beta peptide, the accumulation of which is considered to be a major reason for Alzheimer's disease. This dataset has also been analyzed in Cai and Sun \cite{Ca17}. The compound library consists of 51840 compounds. The dataset consists of three sets of standardized z-scores. The three replications of the z-scores are summarized in a histogram in Figure \ref{figure1_data}.

We apply the Monte Carlo method explained in Appendix \ref{sec_app_1} to estimate the parameters. Before applying the parameter estimation we first de-mean the z-scores. All the estimates of the parameters including the value of the mean is summarized in Table \ref{table_par}. We note that the estimates for the z-scores B and C are close to that obtained in Cai and Sun \cite{Ca17}. The estimate of the non-null proportion for z-score A is higher than the others. This may be explained by the heavier left-tail as observed in the histogram. For the rest of our analyses, we choose to work with the parameter estimates obtained from z-scores B.

\begin{table}[!h]\caption{Estimates of parameters for the data study.} \footnotesize
\
\\
\renewcommand{\arraystretch}{1}\centering
\label{table_par}
\begin{tabular}{c c c c c}\hline\hline
$Parameters$ & Cai and Sun, 2015 & z-scores A & z-scores B & z-scores C\\
\hline
$Mean$& 0.257 & 0.267 & 0.287 & 0.261\\
$p$& 0.0087 & 0.0803 & 0.0132  & 0.0165\\
$\sigma_0^2$& 0.5776 & 0.5240 & 0.5677 & 0.5027\\
$\sigma_{\mu}^2$& -- & 1.7044 & 3.0735  & 2.6817\\
\hline\end{tabular}\
\end{table}

The information on costs for the stages I and II screening is not available for this study. For illustration purpose, we assume that the total budget for the study is $B = \$500,000$. The cost for the stage I screening is fixed at $c_1 = \$1$ and that for stage II is varied over different vales from low to high. The precision of the stage II experiment, relative to stage I experiment, is chosen to be 3. The proposed algorithm is then implemented to determine the optimal number of replications for the stages, and the number of compounds selected for stage II screening. The library for stage I screening consists of $m_1 = 51840$ compounds as before. 

The results from the proposed algorithm are summarized in Table \ref{Table_optimal}. For a fixed precision it is observed that the number of hits identified at stage I decreases with increase in cost of stage II. This is not counterintuitive as it allows for more replications at stage II due to the fixed budget constraint. Moreover we observe that, the overall expected power, again at the optimal selection, decreases. This may be explained by the fact that the procedure automatically adjusts to control the false discovery rate at the desired level, which results in the procedure having lower power for more expensive stage II experiments, everything else being fixed. What is further interesting to note is, the number of optimal replications in stage I does not change very much with change in cost of stage II. One explanation could be that: a reasonable number of replications is essential to maintain the proportion of true compounds identified at stage I.

\begin{table}[!h]\caption{Optimal combination and expected values. Notation $\pmb \theta := (\theta_1, \cdots, \theta_m)$.}\footnotesize \label{Table_optimal}
\
\\
\renewcommand{\arraystretch}{1}\centering\begin{tabular}{c c c c c c}\hline\hline
$c_2 \mbox{ in } \$ $ & $r_1$ & $|A_1|$ & $r_2$ & $E(|A_2|)$ & $E(|A_2 \cdot \pmb \theta|)$\\
\hline
$2$ & 7 & 4936 & 13 & 558.60 & 523.95\\
$5$ & 7 & 3036 & 10  & 541.01 & 505.46\\
$10$ & 7 & 1836 & 7 & 521.54  & 486.11\\
$50$ & 6 & 1136 & 3  & 475.35 & 443.60\\
\hline\end{tabular}\
\end{table}

\section{Discussion} \label{sec_disc}

In this article we developed a two-stage computational procedure for optimal design and error control for HTS studies. By utilizing the Monte Carlo based simulation techniques, our data-driven design calculates the optimal replicates at each stage. The new design promises to significantly increase the signal to noise ratio in HTS data. This would effectively reduce the number of false negative findings and help identify useful compounds for drug development in a more cost-effective way. By controlling the FDR at the confirmatory screening stage, the false positive findings and hence the financial burdens on the hits follow-up stage can be effectively controlled. Finally, under our computational framework, the funding has been utilized efficiently with an optimized design, which allocates available budgets dynamically according to the estimated signal strengths and expected number of true discoveries.

We have assumed a two-point normal mixture model. Although this is a reasonable assumption in some applications, it is desirable to extend the theory and methodology to handle situations where the data follow skewed normal, or skewed t distributions (see, for example, Fr\"{u}hwirth-Schnatter and Pyne \cite{Fr10}). In addition, the two-point mixture model can be extended to a k-point normal mixture model where k may be unknown. The problem for estimating k and the mixture densities has been considered, for example, by Richardson and Green \cite{Ri97}. 

\section*{Acknowledgements}
We thank the Associate Editor and referees for several suggestions that greatly helped to improve both the content and the presentation of this work.

\appendix

\section{Appendix} \label{sec_app}
Here we provide more details for the estimation of the model parameters and implementation of the proposed algorithm.


\subsection{Monte Carlo estimates of parameters} \label{sec_app_1}
For the two-component random mixture model, 
\[
(x_i |\theta_i, \mu_i) \sim (1 - \theta_i) f_0 + \theta_i f_{1i},
\]
assume that $f_0 \sim N (0, \sigma_0^2)$, and $f_{1i} \sim N (\mu_i, \sigma_0^2)$. Each $\theta_i$ follows independent $\mbox{Ber}(p)$. Further let $(\mu_i | \theta_i = 1) \sim N(0, \sigma_{\mu}^2)$. Then note that marginally,
\[
x_i \sim (1 - p) N(0, \sigma_0^2) + p N(0, \sigma_1^2),
\]
where $\sigma_1^2 = \sigma_0^2 + \sigma_{\mu}^2$. We choose to assume the mean to be zero for the signals, that is, $N(0, \sigma_{\mu}^2)$. If not, the data has to be appropriately transformed.

Bayesian formulation is used to estimate the parameters $(p, \sigma_0^2, \sigma_{\mu}^2)$. Let $\pi(\cdot)$ denote prior distributions. To define priors for $\pi(p)$ and $\pi(\sigma_0^2, \sigma_{\mu}^2)$ we follow the formulation defined in Scott and Berger \cite{Sc06}, we let
\[
\pi(\sigma_0^2, \sigma_{\mu}^2) \propto (\sigma_0^2 + \sigma_{\mu}^2)^{-2},
\]
and 
\[
\pi(p) \propto p^{\alpha},
\]
where we choose $\alpha = 5.58$ which may be varied. Note that there are other ways of choosing priors for the parameters (for e.g., choosing inverse gamma for the variance, and beta prior for the the non-null proportion). As discussed in Scott and Berger \cite{Sc06} there is no one ``right'' way to do this and can be viewed as a practitioner's discretion. We have provided one formulation that computes estimate of the parameters using preliminary data. Maximum likelihood estimation, computed by expectation-minimization (EM) algorithm, may also be used in this set-up.

To compute the posterior expectations of $\sigma_{\mu}^2$, $\sigma_0^2$, and $p$ we compute
\[
\int_0^1 \int_0^{\infty} \int_0^{\infty} h(\sigma_{\mu}^2, \sigma_0^2, p) \pi (\sigma_{\mu}^2, \sigma_0^2, p| \x) d\sigma_{\mu}^2 d \sigma_0^2 dp,
\]
with $h(\sigma_{\mu}^2, \sigma_0^2, p) = \sigma_{\mu}^2$, $\sigma_0^2$, and $p$ and $\pi (\sigma_{\mu}^2, \sigma_0^2, p| \x)$ denotes the posterior distribution. These integrals are computed using Monte Carlo based importance sampling method. We follow the method presented in Scott and Berger \cite{Sc06}. Section 3 of the reference provides more details. The choice of $a$ in their method is again at user's discretion. Here we choose $a = 5$. Our simulation results confirm the successful estimation of the model parameters.

\newpage

\subsection{Implementation of the proposed procedure} \label{sec_app_2}

Let $(\widehat{p}, \widehat{\sigma}_0^2, \widehat{\sigma}_{\mu}^2)$ be the estimated parameters. The Lfdr statistic for $r$ replications is given by:
\[
\mbox{Lfdr}(\cdot) = \frac{(1 - \widehat{p})N(0, \widehat{\sigma}_0^2/r)}{(1 - \widehat{p})N(0, \widehat{\sigma}_0^2/r) + \widehat{p}N(0, \widehat{\sigma}_{\mu}^2 + \widehat{\sigma}_0^2/r)}.
\]

\begin{algorithm}[H]
 \KwData{
The number of compounds in stage I ($m_1$), costs in stages I and II ($c_1$ and $c_2$ respectively), study budget ($B$), and the parameters for stages I and II $(\widehat{p}, \widehat{\sigma}_0^2, \widehat{\sigma}_{\mu}^2)$ estimated from preliminary data.
 }
 \KwResult{Optimal values of $r_1$ and $|A_1|$.}
 Initialization: set $r_1 = 1$, $|A_1| = 1$, and \textit{optimal expected hits} $= 0$.\

 \While{$r_1 \leq  B/c_1m_1$ and $|A_1| \leq \min\{(B - c_1r_1m_1)/c_2, m_1\}$} 
 {
 \vspace{+1em}
  Simulate 100 observations of stages I and II expressions; \\
  Rank the stage I expressions by the Lfdr values (increasing) and select top $|A_1|$ compounds for stage II expressions;\\
  Find the number of confirmed hits after stage II using (\ref{lfdr proc}) in each observation;\\
  Average over the 100 observations to compute \textit{expected hits}.\\
  \eIf{$\mbox{expected hits} > \mbox{optimal expected hits}$}{
   Set \textit{optimal expected hits} = \textit{expected hits}\;
   Note the optimal values of $r_1$ and $|A_1|$\;
   Increase values of $r_1$ and $|A_1|$.\
   }{
   Increase values of $r_1$ and $|A_1|$.\
  }
 }
 \vspace{+1em}
 \caption{is proposed to determine the optimal values of $r_1$ and $|A_1|$. The value of $r_2$ is then determined by the study budget.}
\end{algorithm}

\end{document}